\documentclass{Interspeech2024}
\usepackage{multirow}

\interspeechcameraready

\title{Self-Supervised Models for Phoneme Recognition: Applications in Children's Speech for Reading Learning}

\name[affiliation={1,2}]{Lucas}{Block Medin}
\name[affiliation={2}]{Thomas}{Pellegrini}
\name[affiliation={1,2}]{Lucile}{Gelin}

\address{
  $^1$Lalilo by Renaissance Learning, 236 rue du faubourg Saint Martin, 75010 Paris, France\\
  $^2$IRIT, Université de Toulouse, CNRS, Toulouse INP, UT3, Toulouse, France}
\email{lucas.block@renaissance.com, thomas.pellegrini@irit.fr, lucile.gelin@renaissance.com}


\begin{document}

\maketitle

\begin{abstract}
    
    Child speech recognition is still an underdeveloped area of research due to the lack of data (especially on non-English languages) and the specific difficulties of this task. Having explored various architectures for child speech recognition in previous work, in this article we tackle recent self-supervised models. We first compare wav2vec 2.0, HuBERT and WavLM models adapted to phoneme recognition in French child speech, and continue our experiments with the best of them, WavLM base+. We then further adapt it by unfreezing its transformer blocks during fine-tuning on child speech, which greatly improves its performance and makes it significantly outperform our base model, a Transformer+CTC. Finally, we study in detail the behaviour of these two models under the real conditions of our application, and show that WavLM base+ is more robust to various reading tasks and noise levels.
\end{abstract}

\section{Introduction}

Reading tutors have a significant pedagogical impact on children learning to read, and several initiatives have been developed over time~\cite{Mostow2001-ETT, Bolanos11-FOR,Godde2017-EOR}. 
%Anonymisation : Lalilo\footnote{\url{https://www.lalilo.com/}} 
We have created a reading assistant for children aged 5 to 8, including a read-aloud exercise that provides personalised feedback thanks to the automatic phoneme recognition system presented in this article.

The oral language of young children (5-8 years old) has specific characteristics linked to the development of their vocal apparatus and motor control capacities: unstable articulatory mechanisms and intra- and inter-speaker spectral variability~\cite{Lee99-AOC}, fundamental and higher formant frequencies~\cite{Mugitani12-DOV}, and the presence of phonological errors~\cite{Fringi15-EOP}. These morphological and phonological differences, as well as the lack of child speech data, are the main reasons for the limited performance of automatic speech recognition (ASR) systems for children~\cite{Potamianos2003-RRO,Shivakumar20-TLF,Yeung18-OTD}.

%Previous research on children's ASR has shown inferior performance to that observed in adults. 
Hybrid deep neural network - Hidden Markov Model systems obtained improvements by combining adult and child data~\cite{Serizel14-DNN}, or by using transfer learning techniques~\cite{Shivakumar20-TLF}. Supervised end-to-end architectures have recently been adapted to child ASR, and have reached or surpassed the performance of hybrid architectures~\cite{Shivakumar2021-E2E,Gelin2022-JEP}.
Recently, self-supervised learning (SSL) has been introduced into the field of ASR because of its great potential to improve low-resource tasks by exploiting prior knowledge acquired from large amounts of unlabeled data~\cite{Mohamed2022-SSspeech}.
%,Krishna2021-ULS}. 
This is the case in child ASR, where data is scarce and annotation is complex and expensive. Recent studies have shown that the learning potential from abundant unlabeled data is high for child ASR~\cite{Jain2023-AWB,Fan2022-TBD,Li2024-ASSL}. 

Our educational use case however demonstrates specific difficulties, such as young age (5-8), unusual tasks (pseudoword reading), non-English language and classroom noise. The impact of SSL models has not yet been studied for child speech with those specificities. 
Our contributions on this paper are therefore multiple:
\begin{itemize}
    \item We compare the performance of state-of-the-art SSL models (wav2vec 2.0, HuBERT, WavLM), and show the adaptability of these adult-trained English model for phoneme recognition on young children's speech in another language (French) ;
% pretrained in English, for phoneme recognition on young children's speech in French, and thus show the adaptability of an adult-trained English model for young children's speech in another language;
    \item We explore several finetuning processes for adapting the new WavLM model to our application with a small quantity of French children's speech data, and significantly outperform our baseline model, a supervised Transformer+CTC;
    \item We demonstrate that the WavLM base+ model displays better generalization capabilities and better robustness to noise than our supervised model.
\end{itemize}

\section{Speech material}

\subsection{Adult speech}

For our baseline model, we use a version of the French Common Voice\footnote{\scriptsize Corpus available on : \url{https://voice.mozilla.org/fr}} which contains around 150 hours of read speech. The self-supervised models are trained and finetuned on adult speech from the following corpora:
\begin{itemize}
    \item Librispeech~\cite{Panayotov2015-LiSp}: 960 hours transcribed English read speech;
    \item LibriLight~\cite{Kahn_2020}: 60k hours unlabeled English read speech;
    \item VoxPopuli~\cite{wang2021voxpopuli}: 24k hours unlabeled English speech;
    \item GigaSpeech~\cite{chen2021gigaspeech}: 10k hours unlabeled English read and spontaneous speech.
\end{itemize}

\subsection{Child speech}
\subsubsection{In-house child speech corpus in French} %Anonymisation : Lalilo
\label{sec:lalilo-data}
%Anonymisation : The Lalilo
Our in-house (IH) corpus contains recordings of French children from 1st to 3rd grade (age 5- 8), reading aloud various types of content. The data is transcribed manually at word level and each word is labelled  "correct" or "incorrect". Correct words are automatically phonetized, while incorrect words are manually transcribed at the phoneme level. Annotation is done by two annotators, and the recording is discarded when they disagree.

When learning to read, pupils perform various reading tasks, requiring them to use different cognitive processes. 
%Anonymisation :In the Lalilo 
In our reading platform, we offer four main types of content, with varying degrees of difficulty: isolated words, short sentences, word lists and pseudoword lists.
The recordings are mainly collected as part of the oral reading exercise on the platform, which is most often used in classrooms under reduced supervision: they contain varying levels of babble noise. The noise level is calculated using a signal-to-noise ratio (SNR).

The training and validation sets respectively contain 13 hours and 25 minutes of data. Having been designed before the addition of new types of content, these sets contain only isolated words and sentences. In addition, they are composed solely of correctly read utterances. The transcription corresponds to the text requested from the student, automatically phonetized using a pronunciation dictionary. The training and validation sets have mean SNRs of $21.0\pm13.0$ dB and $20.6\pm12.6$ dB respectively.
%, with standard deviations of 13.0 dB and 12.6 dB. 
The test set consists of 3 hours of utterances, with approximately 25\% of the words containing a reading error. We use all four content types here, dividing the test set into subcategories: isolated words (W, 51 min), sentences (S, 29 min), word lists (WL, 56 min) and pseudoword lists (PWL, 50 min). The test's SNR values are identical to the validation set.

\subsubsection{MyST}

The My Science Tutor (MyST) Children's Speech Corpus ~\cite{pradhan2023science} is a large-scale collection of spontaneous American English conversational speech between 3rd, 4th, and 5th grade students and a virtual science tutor. 
%The corpus consists of 393 hours of speech from 1,371 students, comprising 228,874 utterances across 10,496 sessions. The scientific content covered in the MyST corpus is aligned with the Full Option Science System (FOSS) modules, which are used in over 100,000 classrooms across the United States.
Approximately 45\% of all utterances, amounting to 224 hours of speech, have been transcribed at the word level and are presented in .trn file format. 
%Despite the presence of noise in the data and some necessary cleaning steps, the MyST corpus remains the largest open-source dataset of spontaneous child speech available for research purposes, making it a valuable resource for recent studies in the field.

% To prepare the MyST corpus for phonetization, several data cleaning steps were performed. First, all utterances containing typographical transcription errors or non-word labels such as DISCARD, SILENCE, and NO\_SIGNAL were removed. Secondly, filled pauses, non-speech events, truncated words, and unintelligible words were deleted from the transcriptions. %Finally, utterances that could not be phonetized, primarily due to transcription typos, were removed.

To prepare the MyST corpus for phonetization, several data cleaning steps were performed. First, all utterances containing typographical transcription errors or non-word labels such as DISCARD, SILENCE, and NO\_SIGNAL were removed. Secondly, filled pauses, non-speech events, truncated words, and unintelligible words were deleted from the transcriptions. After cleaning, we ended up with a train set of 161 hours, a development set of 25 hours, and a test set of 27 hours. We used the splits provided by MyST for these datasets.

\section{Systems description}

This section presents the different systems we will be studying in this paper.  We train our systems to recognise phonemes, not words, so that we can detect reading errors more effectively. All our systems are trained with SpeechBrain~\cite{Ravanelli2021-SB}.

\subsection{Baseline system: Transformer+CTC with F-bank}

Proposed by~\cite{Vaswani2017-AIA} and adapted to automatic speech recognition (ASR) by ~\cite{Dong2018-STR}, the Transformer model follows a sequence-to-sequence encoder-decoder end-to-end architecture. 
It is based solely on attention mechanisms and compensates the lack of recurrence with positional encodings, multi-head self-attention and cross-attention modules, and position-wise feed-forward neural networks.
%It is based solely on attention mechanisms, abandoning the recurrent neural networks that are usual in \textit{seq2seq} systems. Recurrence, which is essential for extracting positional information from audio frames, is replaced by positional encodings, multi-head self-attention and cross-attention modules, and position-wise feed-forward neural networks. 
The Transformer+CTC model is complemented by a CTC (\textit{Connectionnist Temporal Classification}) function at the encoder output, which improves performance through multi-objective training (cross-entropy and CTC) and joint attention/CTC decoding~\cite{Watanabe2017-HCA, Karita2019-ITB}.

The choice of this architecture is based on its excellent performance in adult speech recognition tasks~\cite{Karita2019-ACS}, which was confirmed on the speech of children learning to read in our previous work~\cite{Gelin2021-SPECOM, Gelin2022-JEP}. Our model uses Mel F-bank features as input, and follows the architecture used in the papers cited above. 
%but a new version has been trained with SpeechBrain to facilitate comparison with other models. 
%It contains 14.3 million parameters.
Our Transformer+CTC model is trained in two stages: firstly trained on adult speech from the Common Voice corpus, then adapted with transfer learning with the IH child speech corpus.
%Anonymisation :Lalilo children's speech data
All the layers are re-trained during this second stage, as advised by~\cite{Shivakumar20-TLF} for very young children.

\subsection{Self-supervised models}
%Since the introduction of wav2vec~\cite{Schneider2019-wav}, s
%Self-supervised models have become an established part of the ASR field. 
By using unlabeled data to extract latent representations, self-supervised models can achieve state-of-the-art results with up to 100 times less annotated data than other supervised models.
These results are particularly noteworthy in the context of children's speech recognition, where models based on the wav2vec 2.0~\cite{baevski2020-wav2vec} architecture achieve performance similar to that of state-of-the-art supervised models~\cite{Jain2023-AWB}.
For our study, we selected the most widely used self-supervised pre-trained models for ASR: wav2vec 2.0, HuBERT and WavLM.

\subsubsection{wav2vec 2.0}
wav2vec 2.0~\cite{baevski2020-wav2vec} is a self-supervised end-to-end architecture based on convolutional and transformer neural networks. The architecture can be divided into three main parts: an encoder, a contextual transformer network, and a quantization module.

The encoder consists of seven temporal convolution blocks, followed by an activation normalisation layer and a GELU activation function. It replaces the absolute positional encoding with a convolution layer, which acts as a relative positional encoding. This encoding is fed through a GELU function, then concatenated with the encoder outputs, and the whole undergoes a layer normalisation. 
%The network is made up of 12 of these blocks, of dimension 768, with an internal dimension of 3082, and 8 attention heads per block.
Finally, the quantization module also takes the encoder output and transforms it into a set of discrete representations via product quantization.

The wav2vec 2.0 model is pre-trained for a masked prediction task: it aims at predicting the correct quantized latent audio representation in context of an utterance despite the application of a mask on part of the audio frames.
The overall training objective is to minimise the contrast and diversity loss functions.

We use a wav2vec 2.0 Base model\footnote{\scriptsize\url{https://huggingface.co/facebook/wav2vec2-base-960h}}. The model is trained using the standard LibriSpeech 960h dataset~\cite{Panayotov2015-LiSp}.

\subsubsection{HuBERT}

The HuBERT~\cite{hsu2021hubert} model uses the same architecture as wav2vec 2.0, but replaces the quantization module with a K-Means quantization, which involves three fundamental changes:
\begin{itemize}
    \item Discrete representations of speech segments are obtained by discovering hidden units, by assigning a cluster to each audio extract using a K-Means algorithm;
    \item The extraction of representations is iterative, using first the results of an MFCC, then the embeddings of the intermediate layers of the pre-trained model;
    \item The contrast loss and diversity loss functions are replaced by a cross-entropy loss, which simplifies training.
\end{itemize}

In this study, we use a pre-trained HuBERT Base acoustic model\footnote{\scriptsize\url{https://huggingface.co/facebook/hubert-base-ls960}}, also trained on the standard Librispeech 960h.

\subsubsection{WavLM}

The WavLM~\cite{Chen2022-wavlm} architecture follows HuBERT's, while introducing a gated relative position bias into the attention mechanisms. Instead of relying solely on the absolute positions of the key and query vectors, the model considers the relative positions between these vectors when calculating attention scores.

The WavLM model also includes modifications in the pre-training phase. The masked prediction task is replaced by a masked denoising and prediction task. This process, which aims at making the model more robust, involves simulating noisy inputs or overlapping speech, then predicting pseudo-labels of the original audio over the masked region.
We will study two WavLM models:
\begin{itemize}
    \item A WavLM Base model\footnote{\scriptsize\url{https://huggingface.co/microsoft/wavlm-base}}, trained with the same data as the previous models;
    \item A WavLM Base+ model\footnote{\scriptsize\url{https://huggingface.co/microsoft/wavlm-base-plus}}, which has the same architecture as WavLM Base, but is trained on a much larger corpus consisting of LibriLight, GigaSpeech and VoxPopuli data, for a total of around 94,000 hours. This extended corpus makes it possible to improve the performance and robustness of the WavLM model while maintaining a reasonable model size~\cite{Chen2022-wavlm}.
\end{itemize}

\section{Model adaptation and evaluation for child speech phoneme transcription}

We decided to focus on pre-trained SSL models in the \textit{Base} format rather than \textit{Large}. On the one hand, the computing capacity required to train and deploy the \textit{Base} models is much lower due to their smaller number of parameters (95M versus 317M). On the other hand, we can see that, in the case of child speech, the performance improvement is small at the cost of a significant increase in the number of parameters~\cite{Jain2023-AWB}. 
The French pre-trained models were poorly documented and very heterogeneous in terms of the data used, which made the comparison complex, and led us to use English models.

To adapt the SSL systems to our task, we feed the output of the Transformer network (size 768) in a linear projection for phoneme classification.
This layer is composed of 35 classes representing the French phonemes and the ``empty" phoneme. The model is trained in a supervised manner with the IH corpus,
%Anonymisation :Lalilo data, 
with the aim of minimising the CTC loss function (\textit{Connectionist Temporal Classification}).
Phoneme error rate (PER) is used to measure the performance of our systems on this task.

%In this section, we compare the different models presented previously and adapted to the phonetic transcription of children's speech. We also explore two adaptation methods by freezing one or more parts of the models, experiment using other child speech data, and finally analyse the performance of the systems when confronted to our application's specificities.

\subsection{Comparing SSL models for our application}
\label{sec:exp1}
% \begin{table}[htbp]
%   \caption{PER (\%) obtained with different models fine-tuned with Lalilo data on the last CTC layer, decoded with a greedy search}
%   \label{tab:comparison}
%   \centering
%   \begin{tabular}{c c c}
%     \toprule
%     \textbf{Model} & \textbf{Lalilo-EN} & \textbf{Lalilo-FR} \\
%     \midrule
% 	wav2vec 2.0 & 62.9 & 62.9 \\
% 	HuBERT &  58.7 & 46.3\\
%     WavLM base & ? & 46.8 \\
%     WavLM base+ & \textbf{45.2} & \textbf{41.5}\\
%     \bottomrule
%   \end{tabular}
% \end{table}

Our first objective is to compare available SSL models for our application. We adapt the models by re-training only the CTC phoneme classification layer with the IH corpus.
%Anonymisation :Lalilo data.
The models are trained on 30 epochs, on AWS g4dn.xlarge GPUs for 10 hours, which consumes 325 gCO2eq\footnote{\scriptsize\url{https://engineering.teads.com/sustainability/carbon-footprint-estimator-for-aws-instances/}} per training. The checkpoint with the best PER on the IH validation set is retained. For this preliminary experiment, we decode with a greedy search.

\vspace{-0.2cm}
\begin{table}[htbp]
  \caption{PER obtained on IH test set with different models with the CTC layer fine-tuned on the IH corpus (greedy search)}
  \label{tab:comparison}
  \centering
  \begin{tabular}{|l|c|}
    \hline
    \textbf{Model} & \textbf{PER (\%)} \\
    \hline
	wav2vec 2.0  & 62.9 \\
	HuBERT & 46.3\\
    WavLM base & 46.8 \\
    WavLM base+ & \textbf{41.5}\\
    \hline
  \end{tabular}
  \vspace{-0.3cm}
\end{table}

% We can see that the performance of the HuBERT and WavLM models far exceeds that of the wav2vec model. For the same amount of data, the difference between HuBERT and WavLM base is not significant. However, the PER obtained by the WavLM base+ model, which has the same number of parameters but is trained on 100 times more data, is significantly better. The rest of the study will therefore focus on this model.

The HuBERT and WavLM models significantly outperform wav2vec 2.0, likely due to their use of K-means clustering for quantization. This approach improves generalization by learning discrete representations that capture high-level acoustic patterns from the unlabeled data. While HuBERT and WavLM base show comparable performance when trained on the same amount of data, the WavLM base+ model achieves a substantially lower PER. This improvement can be attributed to its pretraining on 100 times more data, which allows the K-means clustering to discover more robust and generalizable representations.
%Additionally, WavLM base+ has encountered the French language during its pretraining on the multilingual VoxPopuli dataset, which can account for some improvement. 
The increased generalization ability makes WavLM base+ a promising candidate for further exploration in this study.

\subsection{Adapting WavLM base+ model with in-house data}

We go further in adapting the WavLM model to improve its performance on our application. Rather than just training the CTC layer with child speech, we also adapt part of the pre-trained model. To do this, we follow what is done in~\cite{Jain2023-AWB} for adapting a wav2vec 2.0 model to children's speech: for the first 1000 iterations, only the last CTC classification layer is trained, then the Transformer block is also trained. The CNN encoder, on the other hand, remains frozen. The learning rate is set to 5e-4 and the batch size to 128, following recommendations in~\cite{Chen2022-wavlm}. The model is trained on 55 epochs on an AWS g4dn.12xlarge GPUs (8.5 hours, 1.96 kgCO2eq).

Table~\ref{tab:beam-results} displays the PER values obtained by the baseline Transformer+CTC model, and two WavLM models: the one in the~\ref{sec:exp1} section where only the CTC layer has been adapted and the one adapted at greater depth, respectively denominated as ``frozen" and ``full". Here and in the following, the decoding used for all the models is a beam search (size 10).

\vspace{-0.2cm}
\begin{table}[htbp]
  \caption{PER obtained on IH test set with Transformer+CTC and WavLM base+ models fine-tuned on IH data (beam search)}
  \label{tab:beam-results}
  \centering
  \begin{tabular}{|l|c|c|}
    \hline
    \multirow{2}{*}{\textbf{Model}} & \textbf{\# train. params} & \multirow{2}{*}{\textbf{PER (\%)}} \\
    & \textbf{(\# total)} & \\
    \hline
	Transformer+CTC & 14 M (14 M) & 40.5 \\
    \hline
    WavLM base+ IH-frozen & 28 k \hspace{4pt} (95 M) & 39.2 \\
    WavLM base+ IH-full & 90 M (95 M) & \textbf{26.1}\\
    \hline
  \end{tabular}
\end{table}
\vspace{-0.2cm}
We first observe that the WavLM base+ IH-frozen model achieves a slightly better performance than the Transformer+CTC model, while only its phoneme classification layer (less than 1‰ of the model weights) has been trained with child speech. This shows that the self-supervised representations of the WavLM base+ adult model, despite not having seen child speech when trained, are generic and easily adaptable to different speech types. Also, WavLM having been trained on English data, we can deduce that the representations well adapt to other languages. However, these results must be taken with precautions: the two models obtain comparable results, but the Transformer+CTC contains almost 7 times less parameters.
Unfreezing the Transformer block of WavLM base+ gives a PER of 26.1\%, a relative reduction of 33.4\% compared with the frozen model. This result shows that WavLM representations can nevertheless be adapted to better match a specific kind of speech, and that this adaptation is effective despite a small amount of adaptation data (13 hours).

% \begin{table*}[htbp]
%   \caption{PER (\%) of diverse models evaluated on MyST and Lalilo test sets, decoded with a beam search}
%   \label{tab:comparison}
%   \centering
%   \begin{tabular}{l c c c c c c}
%     \toprule
%     \textbf{Model} & \multicolumn{2}{c}{\textbf{Fine-tuning}} & \textbf{MyST-eval} & \textbf{Lalilo-EN} & \textbf{Lalilo-FR}\\
%     & 1 & 2& & & \\ 
%     \midrule
%     Transformer+CTC  & Lalilo & & - & 29.8 & 40.5\\
%     \hline
%     %WavLM base+ &  & MyST (frozen) & & & & - \\
%     WavLM base+ Lalilo-1 & Lalilo (frozen) & & - & 45.2 & 39.2 \\
%     WavLM base+ Lalilo-2 & Lalilo (full) & & - & 20.3 & 26.1\\
%     \hline
%     WavLM base+ MyST & MyST (full) & & 11.8 & 57.5 & - \\
%     WavLM base+ MyST-Lalilo-1 & MyST (full) & Lalilo (frozen) & - & & \\
%     WavLM base+ MyST-Lalilo-2 & MyST (full) & Lalilo (full) & - & & 36.3 \\
%     \bottomrule
%   \end{tabular}
% \end{table*}

\subsection{Leveraging other child speech data}
% Separate tables
Having only a small amount of child speech data, we want to explore leveraging larger child speech datasets to improve our models performance. In this section, we ``full" fine-tune our pre-trained adult WavLM base+ model with the MyST train dataset (161 hours). The training lasted 34 hours on AWS g4dn.12xlarge GPUs (7.86 kgCO2eq). We validate this model by testing it on the MyST evaluation corpus, on which it obtains a very good PER of 11.8\%. This result compares to the one obtained in~\cite{Li2024-ASSL} on the MyST corpus with a smaller test set (recordings shorter than 15 seconds). We then apply frozen and full fine-tuning with the IH dataset.

\vspace{-0.2cm}
\begin{table}[!htb]
  \caption{PER obtained on IH test set with WavLM base+ models fine-tuned on MyST then IH data}
  \label{tab:myST-results}
  \centering
  \begin{tabular}{|l|c|}
    \hline
    \textbf{Model} & \textbf{PER (\%)} \\
    \hline
    %WavLM base+ Lalilo-frozen & 39.2\\
    %WavLM base+ Lalilo-full & 26.1\\
    %\hline
    WavLM base+ MyST-IH-frozen & 58.8 \\
    WavLM base+ MyST-IH-full & 36.3 \\    
    \hline
  \end{tabular}
\end{table}
\vspace{-0.2cm}

We can see in Table \ref{tab:myST-results} that using the MyST data does not bring the expected improvements. We can form several hypothesis. First, the two datasets might be too different: MyST students are older (which makes their voice significantly different~\cite{Yeung18-OTD}), speech is spontaneous versus read, and the MyST vocabulary is quite specialized. This would explain that using a model fine-tuned on adult read speech works better than using a model fine-tuned on far-from-domain child speech. Secondly, the models have been first pre-trained and fine-tuned on adult English speech, then fine-tuned on child English speech, to be finally tested on child French speech. Its bad performance could be due to an over-fitting on the English phoneme representations.
%These results emphasize the importance of using data similar to the application. 
The discussion will thus be held on our best model, the WavLM base+ IH-full, that obtains the lowest PER (26.1\%).

\vspace{-0.2cm}
\section{Discussion}

In the previous section, we saw that the Transformer+CTC and our best WavLM base+ model show a PER difference of 14.4\%. In this section, we want to find out whether this difference is evenly distributed across diverse reading tasks and noise levels.

\vspace{-0.2cm}
\subsection{Application to diverse reading tasks}

We now explore the performance of the systems according to the different reading tasks proposed to the students (see section~\ref{sec:lalilo-data}). We easily see in table~\ref{tab:reading-tasks} that the PER difference between the two models does indeed depend on the reading task.

\vspace{-0.1cm}
\begin{table}[!htb]
    \centering
        \caption{PER (\%) obtained with Transformer+CTC and WavLM base+ IH-full models, depending on reading tasks (S = sentence, W = word, WL = words list, PWL = pseudowords list)}
	\begin{tabular}{|l|c c c c|}
    \hline
	\multirow{2}{*}{\textbf{Model}} & \multicolumn{4}{c|}{\textbf{Reading task}} \\
    & S & W & WL & PWL  \\
    \hline
    Transformer+CTC & 16.5 & 34.0 & 46.5 & 59.0  \\
    WavLM base+ IH-full & 16.4 & 25.5 & 28.3 & 32.9 \\
    \hline
	\end{tabular}
    \label{tab:reading-tasks}
\end{table}

\vspace{-0.2cm}
Short sentences (S in the table) represent the easiest task for ASR, with a sufficient but not too large context, and the presence of linking words commonly seen in training. It is also a classic task for adult speech corpora. 
Both models perform similarly, indicating that on an easy and well-known task,
%On this subset, there was no significant difference between the two models. Both models are adapted with the same amount of child speech. On this easy and well-known task, 
supervised learning of a small model (14M parameters) with 150 hours of adult speech is as effective as unsupervised learning of a large model (95M) with almost 100,000 hours of speech.

For phoneme recognition in isolated words (W), WavLM is significantly better (-8.5\% absolute). Since words can contain as few as 2 phonemes, the Transformer+CTC's difficulty can be explained by the lack of context, hurting the attention mechanisms. This phenomenon was notably observed in~\cite{Chan2015-LAS}, where the model's performance degrades significantly when the utterance contains only a single word. WavLM also contains a Transformer block which is affected by this phenomenon, but it is probably compensated for by the use of a CNN encoder, whose convolutions make the most of the lack of context.

The word lists (WL) and pseudoword lists (PWL) were not seen during training, making them slightly out-of-domain. This is even more the case for the pseudoword lists because pseudowords do not exist and have never been seen in any corpus of adult or child speech. WavLM model clearly outperforms the Transformer+CTC on these tasks. We also observe that the more outside the domain is the task, the greater is the relative reduction in PER provided by WavLM: -39\% on word lists, -44\% on pseudoword lists. We can deduce from these observations that the WavLM model has a better generalisation capacity, which is undoubtedly linked to the quantity of data encountered, but also to self-supervised learning, which is less constrained and thus creates more generic representations.

% Possible reformulation:
% Short sentences represent the easiest task, with sufficient context and common linking words. Both models perform similarly, indicating that supervised learning of a small model with 150 hours of adult speech is as effective as unsupervised learning of a large model with almost 100,000 hours of speech for this task.

% For isolated words, WavLM significantly outperforms Transformer+CTC (-8.5\% absolute). The Transformer+CTC's difficulty can be explained by the lack of context, hurting the attention mechanisms. WavLM's CNN encoder likely compensates for this.

% Word lists (WL) and pseudoword lists (PWL) were not seen during training, making them slightly out-of-domain. This is particularly the case for the pseudoword lists be-
% cause pseudowords do not exist and have never been seen in
% any corpus of adult or child speech. WavLM outperforms Transformer+CTC, with the relative PER reduction increasing for more out-of-domain tasks (-39\% on WL, -44\% on PWL). This suggests WavLM has better generalisation capacity, likely due to the quantity of training data and nature of self-supervised learning.

\vspace{-0.2cm}
\subsection{Robustness to classroom noise}
We also want to study the behaviour of our two systems in real classroom conditions. We divide our test set into three noise levels, and compute the PER on each (table~\ref{tab:noise}): low noise (SNR above 25 dB), medium noise (SNR between 10 and 25 dB) and high noise (SNR below 10 dB).
% \begin{itemize}
%     \item Low : recordings with a SNR above 25 dB ;
%     \item Medium : recordings with a SNR between 10 and 25 dB ;
%     \item High : recordings with a SNR below 10 dB.
% \end{itemize}

\vspace{-0.2cm}
\begin{table}[!htb]
    \centering
    \caption{PER (\%) obtained with Transformer+CTC and WavLM base+ IH-full models, depending on the noise level}
	\begin{tabular}{|l|c c c|}
    \hline
	\multirow{2}{*}{\textbf{Model}} & \multicolumn{3}{c|}{\textbf{Noise level}} \\
    & low & medium & high \\
    \hline
    Transformer+CTC & 14.6 & 24.6 & 40.6 \\
    WavLM base+ IH-full & 13.4 & 21.7 & 31.6 \\
    \hline
	\end{tabular}
    \label{tab:noise}
\end{table}
\vspace{-0.2cm}
It is clear that, for both models, performance deteriorates sharply with increasing noise level. Interestingly, the difference in PER between the two models increases with noise level: 1.2\% at low noise, 2.9\% at medium noise and 9.0\% at high noise. The WavLM model is therefore more robust to noise. 

To confirm this observation, we look at the performance of the models as a function of noise level on the sentence test subset, on which the models obtain a comparable PER.
\begin{itemize}
    \item Transformer+CTC : 10.6 (low) - 17.1 (medium) - 30.0 (high)
    \item WavLM base+ \hspace{14pt} : 12.5 (low) - 17.1 (medium) - 26.8 (high)
\end{itemize}
This shows that WavLM is more robust in high noise conditions, at the cost of poorer performance in low noise conditions. These results are in line with the changes introduced in the WavLM pre-training, aimed at making the model more robust to difficult acoustic conditions~\cite{Chen2022-wavlm}.

%\subsection{Robustness to reading mistakes}
\vspace{-0.2cm}
\section{Conclusion}

Accurate transcription of children's speech remains challenging, particularly in French, due to data scarcity and the inherent difficulties of this speech type. In this work, we explored adapting self-supervised models to phoneme recognition in young children's speech. 
When comparing wav2vec 2.0, HuBERT and WavLM \textit{Base} models by training only a CTC phoneme classification layer, we found that HuBERT and WavLM outperformed wav2vec 2.0. The WavLM base+ model, trained on significantly more data, achieved the best performance.
%First, we select three models (wav2vec 2, HuBERT and WavLM, in the \textit{Base} version) and adapt a CTC phoneme classification layer to our small corpus of children's speech. We observe that the HuBERT and WavLM models outperform wav2vec 2, and that the WavLM base+ model, trained on 100 times more data while retaining the same number of parameters, performs significantly better than the others. 
Further adapting WavLM base+ by unfreezing the Transformer blocks during fine-tuning improved its accuracy by 33.4\%, reaching a 26.1\% PER and surpassing our Transformer+CTC baseline.
%Secondly, we adapt the WavLM base+ model in greater depth by unfreezing the Transformer blocks of the model, which improves its performance by a relative 33.4\%. We show that it achieves significantly better accuracy than our base model, a Transformer+CTC. 
Experiments leveraging additional child speech data to overcome data scarcity yielded negative results, potentially due to domain and language mismatches.
%We tentatively leverage other child speech data in the objective of overcoming our lack of data, but obtain worse results, which could be explained by an overfitting on data that is too far from our application, and in a different language.
Analyzing model performance on various tasks and noise conditions, we demonstrated WavLM base+'s superiority on short recordings, better generalization to unseen content, and increased robustness to classroom noise.
%Finally, we analyse the behaviour of our models when faced with different playback tasks and noise conditions, and show that WavLM base+ performs better on very short recordings, generalises better to unseen learning content, and is more robust to classroom noise.

\bibliographystyle{IEEEtran}
\bibliography{mybib}

\end{document}